\documentclass[final,1p,times]{elsarticle} 
\usepackage{graphicx} 
\usepackage{amssymb} 
\usepackage{amsthm} 
\usepackage{lineno} 

\journal{Nuclear Physics A} 
\begin{document} 

\begin{frontmatter} 


\title{Inclusive Dielectron Production in Ar+KCl Collisions at 1.76\,A~GeV studied with HADES}

\author{Filip Krizek$^{a}$ for the HADES collaboration}

\address[a]{Nuclear Physics Institute, \v{R}e\v{z} near Prague, CZ-25068, Czech Republic}

\begin{abstract} 
Results of the HADES measurement of inclusive dielectron production in Ar+KCl collisions at a kinetic beam energy of 1.76\,A~GeV are presented. For the first time, high mass resolution  spectroscopy was performed. The invariant mass spectrum of dielectrons is compared with  predictions of UrQMD and HSD transport codes.
\end{abstract} 

\end{frontmatter} 



\section{Introduction}
\label{sec:Intr}
The High Acceptance Di-Electron Spectrometer (HADES) at GSI \cite{NIM} was designed to investigate di\-electron emission from heavy-ion collisions in the range of kinetic beam energies 1-2\,A~GeV. In these reactions, hadronic matter can be substantially compressed and heated. Maximum densities may reach up to 3 times the normal nuclear density.  This early dense phase lasts for about 10~fm/$c$. At these energies, the fireball is formed mainly by nucleons and baryonic resonances; meson multiplicities and the strangeness content are small.  The main goal of the HADES experiment is to study properties of hadrons inside the hot and dense nuclear medium via their di\-electron decays. Di\-electrons are considered as an ideal probe since they are not  sensitive to final-state interactions. 
 
 Possible in-medium changes can be well examined in case of the light vector mesons $\rho^{0}$ and $\omega$. These particles live only for a~short time and may decay already inside the dense matter. Light vector mesons have small branching ratios to the $e^{+}e^{-}$ decay channel in of the order of $10^{-4}$. Therefore, investigation of these rare decays is a~challenge to the experimental setup and data analysis methods.

\section{Experiment, Data Analysis, and Results}
\label{sec:Exp}
HADES operates at the GSI Helmholtzzentrum f$\ddot{\textrm{u}}$r Schwerionenforschung in Darmstadt. The heavy-ion beams are provided by the SIS18 accelerator. The spectrometer is described in detail in \cite{NIM}. Here, we will only point out that the main detector for electron/positron identification is a~Ring Imaging Cherenkov detector (RICH). Electron/positron identification algorithms further take advantage of measuring the time of flight in a~Time-of-Flight (TOF) wall and registration of electromagnetic showers in a~Pre-Shower detector.  

In the Ar+KCl measurement, a beam of  $^{40}$Ar ions having a~kinetic energy of 1.76\,A~GeV was used. The beam intensity was about $6\times 10^{6}$ particles per spill (10~s). The fourfold-segmented target was made of natural KCl and had an interaction length of 3.05~\%. In this run, the achieved mass resolution at the $\omega$ pole mass was about 3~\%.

The on-line event selection was done in two steps. The first level trigger picked out those reactions where the number of particle hits exceeded 15 in the TOF scintillators. These events were then examined with the second level trigger to find possible single lepton signatures.  The first level trigger enhanced the mean pion multiplicity approximately two times with respect to the minimum bias multiplicity.  This corresponds to a~mean number of participating nucleons around 38.5 and an average charged pion multiplicity of $(3.4\pm0.4)$, see \cite{Bormio_Pavel}.  

Electron/positron identification was done in three parallel analyses. Identification algorithms were based on (i) multi variate analysis, (ii) a~Bayesian approach, and (iii) an~isolation-cut discrimination technique \cite{NIM}.  Electrons and positrons were assembled into pairs. The total number of reconstructed $e^{+}e^{-}$ pairs, $N_{+-}$, can be decomposed as $N_{+-}=S\,+\,CB$, where $S$ denotes the number of signal pairs and $CB$ stands for the number of combinatorial background pairs. To enhance the signal and to reduce $CB$, it is necessary to suppress contributions from photon conversion, from misidentified hadrons, and from tracking fakes.  The main source of gamma photons are $\pi^0 \rightarrow \gamma\gamma$ decays.  Conversion pairs have often small opening angles and they can be effectively rejected with an opening angle cut $\alpha_{e+e-}>9^\circ$. Tracking fakes were removed from the sample by selecting uniquely defined tracks which were well reconstructed with the tracking algorithm. Further, we applied a~cut $0.1<p_{e}<1.1$~GeV/$c$ on single electron/positron momentum. The upper cut reduces the hadron contamination of the identified electron/positron sample, but does not significantly influence the signal yield.

In the low mass region $M_{ee}<400\mathrm{~MeV}/c^{2}$, where the correlated background from the $\pi^{0}$ to two photon decay followed by double conversion contributes, the combinatorial background was estimated using the method based on like-sign $e^{+}e^{+}$ and $e^{-}e^{-}$ pairs emerging in the same event; in this case $CB=2\sqrt{N_{e+e+}N_{e-e-}}$. At larger masses, where the statistics of like-sign pairs is not sufficient, we used a~mixed event background normalized to the like-sign pair yield.

\begin{figure}[ht]
\centering
\includegraphics[width=0.49\textwidth]{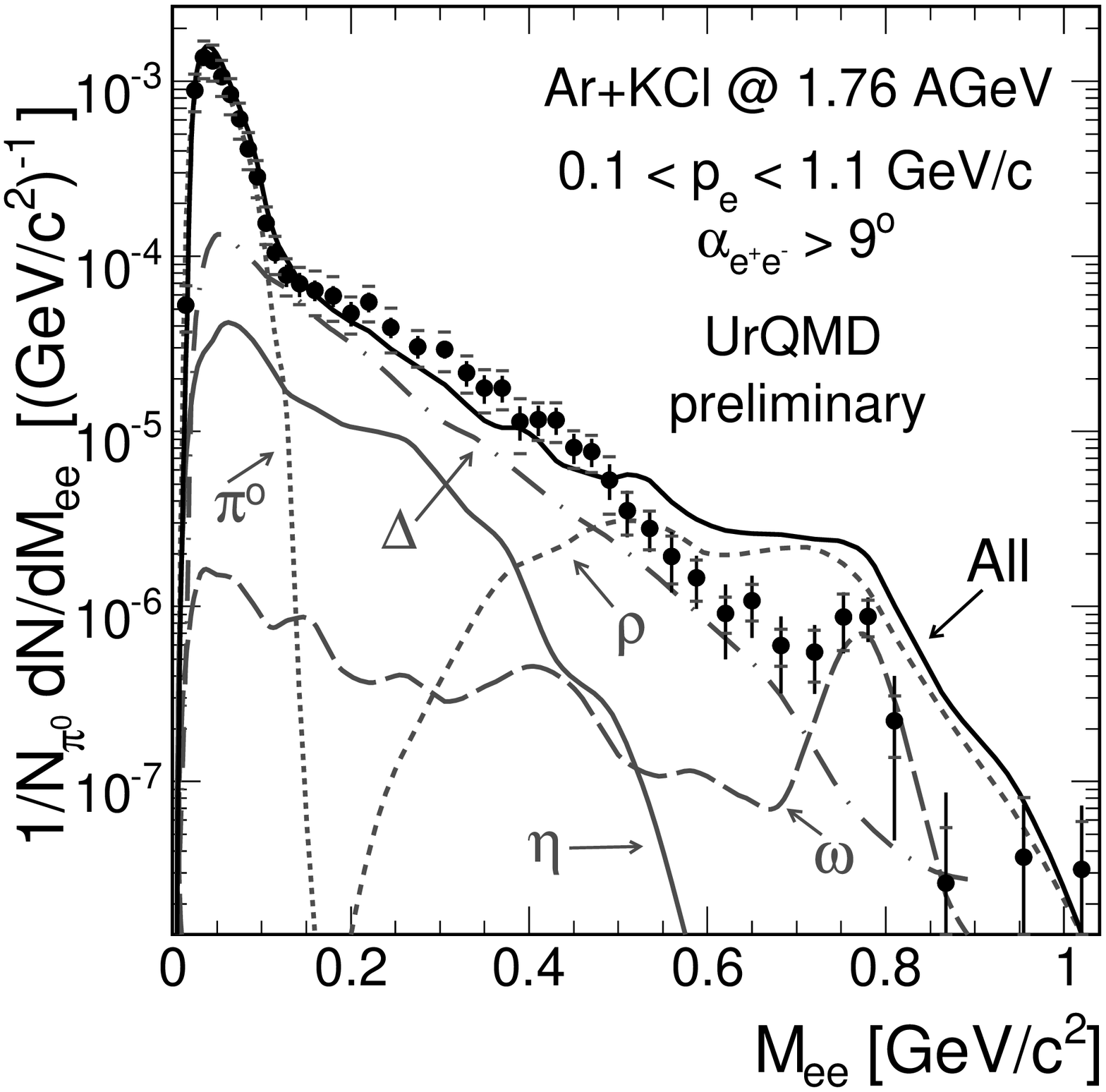}
\includegraphics[width=0.49\textwidth]{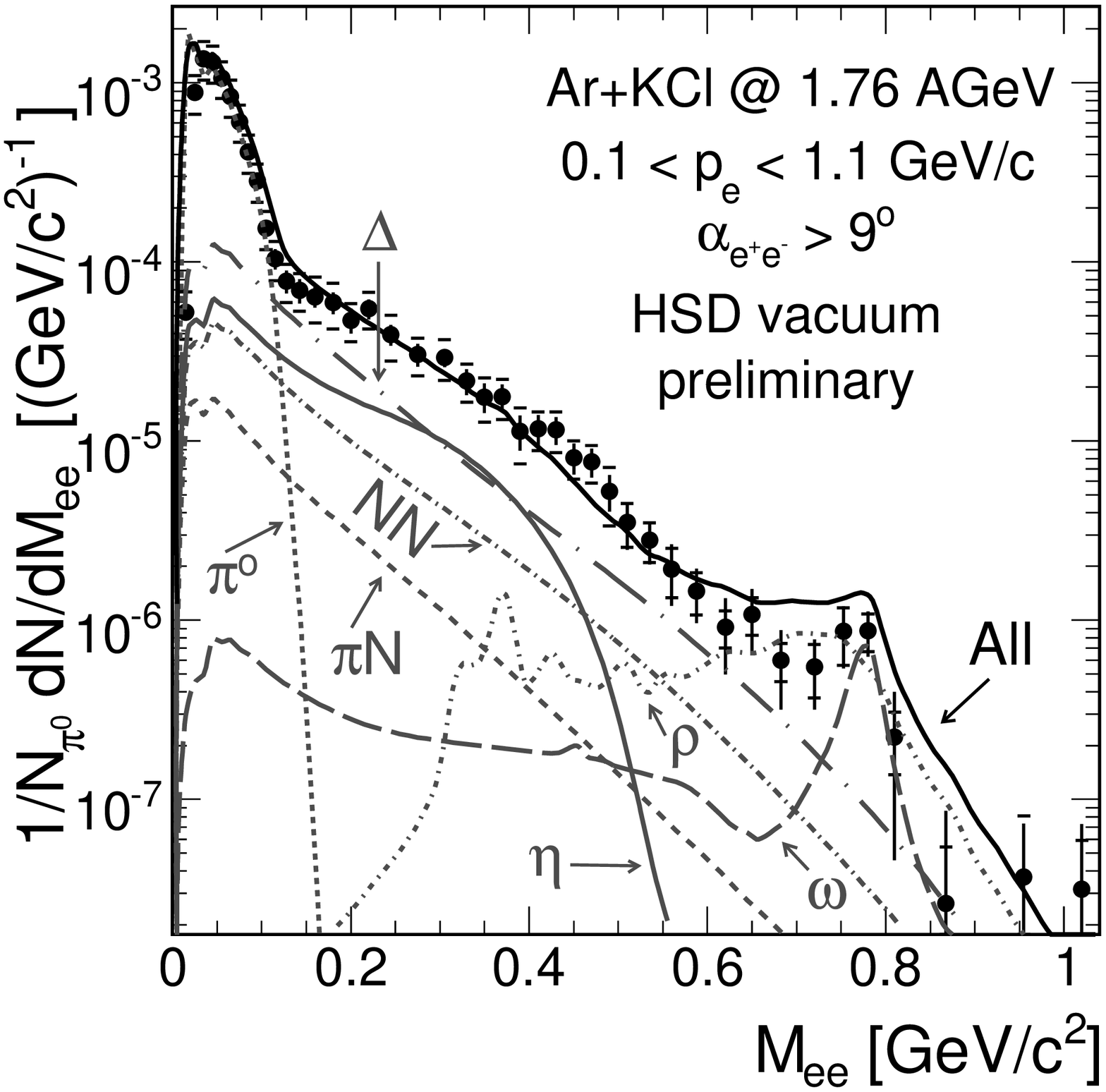}
\caption[]{Invariant mass distribution of $e^{+}e^{-}$ pairs measured by HADES (points) compared with predictions (left: UrQMD \cite{UrQMD}, right: HSD \cite{HSD}; all particles have their vacuum properties; individual contributions are shown too). Statistical and systematic errors of the measurement are shown as vertical and horizontal bars, respectively. }
\label{fig:arkcl}
\end{figure}


The invariant mass spectrum of the di\-electron signal corrected for the detector and reconstruction inefficiencies is shown in Figure~\ref{fig:arkcl}. The spectrum (not acceptance corrected) is normalized to the average number of charged pions assuming $N_{\pi^{0}}=\left(N_{\pi-}+N_{\pi+}\right)/2$. We consider this average as a~good estimate of the $\pi^{0}$ yield. This way of normalization compensates to first order the bias caused by the implicit centrality selection of our trigger. 
 The spectrum shown represents an averaged result from the three parallel analyses mentioned above. Systematic errors are given by the horizontal bars. They stem from the correction on reconstruction efficiency and combinatorial background estimation (20~\%), the uncertainty in our knowledge of the $\pi^{0}$ multiplicity (11~\%), and differences between the three analysis methods. Let us point out that the spectrum exhibits a~clear omega peak in the vector meson region. This is for the first time that vector meson production has been observed in heavy-ion collisions in the SIS/Bevalac energy regime.

In Figure \ref{fig:arkcl}, we contrast our $e^{+}e^{-}$ invariant mass spectrum with predictions by the UrQMD model \cite{UrQMD} and the HSD model \cite{HSD}. Both predictions were generated assuming that hadron properties are not modified by the surrounding nuclear medium, i.e., particles have their vacuum properties. The UrQMD and HSD results were weighted according to our first level trigger impact parameter selection. Both predicted spectra were normalized to their respective $\pi^{0}$ multiplicities.

Both transport codes suggest that below an~invariant mass 0.15~GeV/$c^{2}$, the spectrum is dominated by the di\-electrons from $\pi^{0}$ Dalitz decays. In the intermediate mass region, 0.15--0.5~GeV/$c^{2}$, the pair yield originates mainly from $\eta$ and $\Delta$ Dalitz decays. Dielectron decays of the light vector mesons, $\omega$ and $\rho^{0}$, have substantial contribution to the spectrum only at higher invariant masses.
As compared to the UrQMD calculation, HSD includes additional sources of dielectrons, namely the NN and $\pi$N  bremsstrahlung processes. These contributions to the pair yield are, however, for the Ar+KCl system at given beam energy predicted to be less significant. 

\begin{figure}[ht]
\centering
\includegraphics[width=0.49\textwidth]{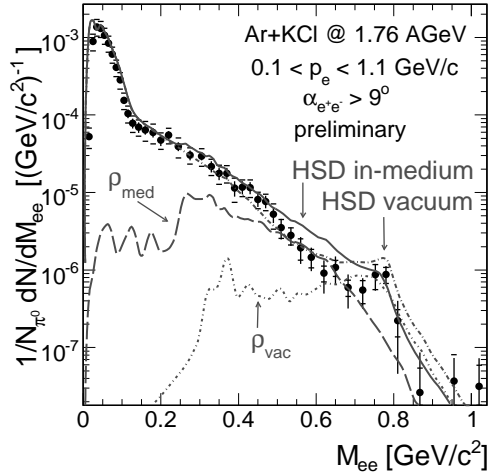}
\caption[]{Invariant mass distribution of $e^{+}e^{-}$ pairs compared with the ``HSD vacuum'' and the `HSD in-medium'' scenarios \cite{HSD}. The ``HSD in-medium'' prediction took into account collisional broadening and simultaneous dropping masses of $\rho^{0}$ and $\omega$ in nuclear environment. Experimental data points are the same as in Fig.~\ref{fig:arkcl}.}
\label{fig:hsd}
\end{figure}

In \cite{HSD}, the HSD group published also a~prediction on how  the di\-electron spectrum would be changed, if $\rho^{0}$ and $\omega$ were modified by simultaneous collisional broadening and mass shifts \`{a} la Brown-Rho \cite{Brown_Rho} in medium. In Figure~\ref{fig:hsd}, we contrast this scenario with the vacuum HSD spectrum and our data. The HSD calculation suggests that the most sensitive part of our spectrum to possible in-medium modifications should be the region of 0.5--0.8~GeV/$c^{2}$.

From the inspection of Figure~\ref{fig:arkcl} and Figure~\ref{fig:hsd} one can conclude that none of the predicted results gives a~fully convincing explanation of the data in the vector meson mass region. One of the reasons might be the parametrization of the elementary cross sections which overestimates the $\rho^{0}$ meson cross section measured in pp collisions \cite{DISTO}.
 

\section{Summary}
We reported on a recent HADES measurement of inclusive dielectron emission from   Ar+KCl collisions at 1.76\,A~GeV. For the first time at SIS energies, the observed di\-electron invariant mass spectrum exhibits a~clear peak in the $\omega$ pole mass region. The presently available HSD and UrQMD transport code predictions overestimate the $\rho^{0}$ production.    

\section*{Acknowledgments} 
The collaboration gratefully acknowledges the support by BMBF grants 06DR135, 06FY171, 06MT238 T5, and 06MT9156 TP5 (Germany), by the DFG EClust 153,  by GSI TM KRUE, by grants MSMT LC07050 and GA ASCR IAA100480803 (Czech Rep.), by grants RII3-CT-2005-515876, FPA2006-09154, and CPAN:CSD2007-00042 (Spain).  

\end{document}